\documentclass[twocolumn,floats,showpacs,superscriptaddress,pre]{revtex4}
\usepackage{amsmath,amssymb,bm}
\usepackage{graphics}
\usepackage{epsfig}
\usepackage{graphicx}

\begin{document}

\title{OBSERVER IN QUANTUM COSMOLOGY}
\author{Natalia Gorobey}
\affiliation{Peter the Great Saint Petersburg Polytechnic University, Polytekhnicheskaya
29, 195251, St. Petersburg, Russia}
\author{Alexander Lukyanenko}
\email{alex.lukyan@mail.ru}
\affiliation{Peter the Great Saint Petersburg Polytechnic University, Polytekhnicheskaya
29, 195251, St. Petersburg, Russia}
\author{A. V. Goltsev}
\affiliation{Ioffe Physical- Technical Institute, Polytekhnicheskaya 26, 195251, St.
Petersburg, Russia}

\begin{abstract}
Within the framework of the new formalism of quantum theory - the quantum
principle of least action - the initial state of the universe is determined,
which is an analogue of the Hartle-Hawking no-boundary wave function. The
quantum evolution of the universe is modified by additional conditions in a
certain compact region of space-time, which is called the observation
region. Additional conditions are Noether identities related to the general
covariance of the theory and internal symmetries of matter fields. The
consequences of the local law of conservation of the energy-momentum tensor
of matter are considered in detail. Its consequence is the deterministic
nature of the motion of the energy and momentum densities of matter in the
observation area. The geometric parameters of the region boundary are also
determined by the deterministic motion of the matter fields inside. The
choice of boundary conditions for the energy-momentum flow at the boundary
serves as a mechanism for decoherence of the quantum evolution of the
universe. The result of decoherence is a certain correspondence between the
final state of the universe and the state of the observer in the specified
region. This correspondence allows us to formulate the extremum principle in
quantum cosmology, in which the action functional constructed using the
final state determines the world history of the universe as the observer
sees it.
\end{abstract}

\maketitle







\section{\textbf{INTRODUCTION}}

The question of observation or measurement in quantum cosmology may seem
far-fetched at a time when the problem of measurements in quantum mechanics
remains unsolved. However, the example of Everett's many-worlds
interpretation of quantum mechanics \cite{Ev} shows us the opposite: perhaps
it is precisely at the cosmological level that we should seek a solution to
the problems of local measurements. The problem is essentially one: how does
the destruction of the quantum superposition of potential possibilities
occur during measurement? Here, of course, preference is given to local
mechanisms of decoherence as a result of the continuous process of
interaction of the quantum system with the environment \cite{Cast,Hack,Mens}%
. But at the same time, the question also arises about the role of the
observer and his consciousness in the process of decoherence \cite{Pen,Davis}%
. What role can quantum cosmology play in this? Any interaction, including
measurement, is reduced to the exchange of energy and momentum (possibly
other physical quantities) between bodies, and this constitutes the content
of the fundamental laws of motion of matter, which are contained in the
covariant law of conservation of the energy-momentum tensor:
\begin{equation}
\nabla _{\mu }T^{\mu \nu }=0  \label{1}
\end{equation}%
It should be emphasized that the tensor conservation law (1) is not simply
part of the equations of motion of matter. It is an identical consequence of
Einstein's equations for geometry, which in turn reflects a deep connection
between symmetries (in this case, general covariance) and conservation laws
(Noether's theorems \cite{No},\cite{KoP}).In the final formulation of the
theory of measurements, one should also take into account the Noether
identities associated with the internal symmetries of matter fields. As for
decoherence, we again refer to Everett's many-worlds interpretation, where
all potential possibilities are realized in different branches of
cosmological evolution, although without specifying a specific mechanism for
this `branching'. In this paper, identity Eq.~(\ref{1}) is proposed as a
mechanism for cosmological decoherence as an additional condition when
quantizing the theory of gravity in some region of space-time. We call such
a region the region of observation or `observer'. The additional condition
Eq.~(\ref{1}) assumes the presence of time in the quantum theory of gravity
(QTG) and the evolution of the universe from some initial state. This is not
present in the `frozen' formalism of QTG, based on the Wheeler-DeWitt
equation \cite{MTW}. Time is also absent in the known scenarios of the birth
of the universe from `nothing' by tunneling \cite{Vil} and the no-boundary
wave function \cite{HH}. In the works \cite{GLG1,GLG2} the problem of time
is discussed in connection with the concept of the proper mass of the
universe, and a formulation of the QTG is proposed, based on the quantum
principle of least action (QPLA). The formalism of the QPLA is equally
suitable for formulating the dynamics and for constructing a theory of the
initial state of the universe in its Euclidean sector. In this paper, we use
a new formulation of the QTG to construct a theory of local observation in
quantum cosmology. Our first task here is to ensure that the proposed
construction has reality in Tegmark's mathematical universe \cite{Teg}.

In Section II we formulate the QPLA, starting with the simplest mechanical
system. In Section III we introduce a generalized canonical form of gravity
theory, in which time and spatial coordinates are equivalent. In Section IV
we formulate the QPLA in the generalized canonical form of gravity theory.
The initial state of the universe within QPLA framework is defined in
Section V. The QTG in generalized canonical form caused by the introduction
of an observation domain in the universe is considered in Section VI.
Section VII formulates a decoherence mechanism in which the observer's state
is one-to-one related to the future state of the universe.

\section{QUANTUM PRINCIPLE OF LEAST ACTION}

Let us formulate the QPLA first for the simplest mechanical system
(particle) with one coordinate $q$, the Hamiltonian function of which is
\begin{equation}
H\left( p,q,t\right) =\frac{p^{2}}{2}+V\left( q,t\right) .  \label{2}
\end{equation}%
The evolution of the particle's state over time is determined by the Schr%
\"{o}dinger equation
\begin{equation}
i\hslash \frac{\partial \psi }{\partial t}=H\left( \widehat{p},q,t\right)
\psi .  \label{3}
\end{equation}%
An alternative formulation of quantum mechanics can be obtained as follows.
Let $\psi (t,q)$ be a solution of the equation Eq.~(\ref{3}) with the
initial condition $\psi (0,q)$. Take an arbitrary trajectory of a particle $%
q=q(t)$ on a time interval $t\in \lbrack 0,T]$ and approximate it by a
broken line with vertices $q_{k}=q\left( t_{k}\right) $, where $%
t_{k}=k\epsilon ,k=1,2,...N$,$\epsilon =T/N$. Consider the product
\begin{equation}
\Psi \left[ q_{k}\right] =\prod\limits_{k}\psi \left( t_{k},q_{k}\right) .
\label{4}
\end{equation}%
This is a function of many variables $q_{k}$. In the limit $N\rightarrow
\infty $ we obtain the wave functional $\Psi \left[ q\left( t\right) \right]
$ defined on the trajectories of the particle $q=q(t)$ in the configuration
space. We introduce the variational derivative of this functional using its
multiplicative approximation Eq.~(\ref{4}) as follows \cite{FeyHib}:
\begin{equation}
\frac{\delta \Psi }{\delta _{t}q\left( t_{k}\right) }=\frac{1}{\epsilon }%
\frac{\partial \Psi }{\partial q_{k}},  \label{5}
\end{equation}%
the notation $\delta _{t}$ indicates that the time function, which is the
argument of $\Psi \left[ q\left( t\right) \right] $, is varied. In the new
formulation of quantum mechanics, we realize the canonical momentum as an
operator of functional differentiation on the space of wave functionals:
\begin{equation}
\widehat{p}\left( t\right) \equiv \frac{\widetilde{\hslash }}{i}\frac{\delta
}{\delta _{t}q\left( t\right) },  \label{6}
\end{equation}%
where, according to Eq.~(\ref{5}),
\begin{equation}
\widetilde{\hslash }=\hslash \epsilon .  \label{7}
\end{equation}%
Substituting Eq.~(\ref{6}) into the canonical form of the particle action,
we obtain the action operator:
\begin{equation}
\widehat{I}_{q}=\int_{0}^{T}dt\left[ \overset{\cdot }{q}\frac{\widetilde{%
\hslash }}{i}\frac{\delta }{\delta _{t}q\left( t\right) }+\frac{\widetilde{%
\hslash }^{2}}{2}\frac{\delta ^{2}}{\delta _{t}q^{2}\left( t\right) }%
-V\left( q,t\right) \right]  \label{8}
\end{equation}%
on the space of wave functionals $\Psi \left[ q\left( t\right) \right] $.
The QPLA is formulated as a secular equation for this operator:
\begin{equation}
\widehat{I}_{q}\Psi =\Lambda \Psi .  \label{9}
\end{equation}%
The assertion is that the multiplicative wave functional Eq.~(\ref{4}) is an
eigenfunction of the action operator if $\psi \left( t_{k},q_{k}\right) $ is
a solution of the Schr\"{o}dinger equation at time $t_{k}$. In this case,
the eigenvalue is:
\begin{equation}
\Lambda =\frac{\hslash }{i}\left[ \ln \psi \left( T,q_{T}\right) -\ln \psi
\left( 0,q_{0}\right) \right] .  \label{10}
\end{equation}%
This is easy to verify by substituting Eq.~(\ref{4}) into Eq.~(\ref{9})
(also replacing the integral with an integral sum), and using the
exponential form of the wave function:
\begin{equation}
\psi \left( t,q\right) =\exp \left[ \frac{i}{\hslash }\chi \left( t,q\right) %
\right] .  \label{11}
\end{equation}%
Note that the wave functional, taking into account Eq.~(\ref{11}), can also
be represented in exponential form in the limit $\epsilon \rightarrow 0$:
\begin{equation}
\Psi \left[ q\left( t\right) \right] =\exp \left[ \frac{i}{\widetilde{%
\hslash }}\int_{0}^{T}dt\chi \left( t,q\left( t\right) \right) \right] .
\label{12}
\end{equation}%
This representation of the wave functional is singular as $\epsilon
\rightarrow 0$. However, it is now completely replaced by the local function
$\chi \left( t,q\left( t\right) \right) $, and the first variational
derivative is defined as:
\begin{equation}
\frac{\widetilde{\hslash }}{i}\frac{\delta \Psi }{\delta _{t}q\left(
t\right) }=\frac{\partial \chi \left( t,q\left( t\right) \right) }{\partial q%
}\Psi .  \label{13}
\end{equation}%
The only difficulty that may arise when substituting Eq.~(\ref{12}) into
Eq.~(\ref{9}) is the calculation of the second variational derivative in Eq.(%
\ref{8}). However, we have:
\begin{eqnarray}  \label{14}
&&\widetilde{\hslash }^{2}\frac{\delta ^{2}\Psi }{\delta _{t}q^{2}\left(
t\right) }  \notag \\
&=&{-}\frac{\widetilde{\hslash }}{i}\frac{\delta }{\delta _{t}q\left(
t\right) }\left[ \frac{\partial \chi \left( t,q\left( t\right) \right) }{%
\partial q}\Psi \right] {-}\left( \frac{\partial \chi \left( t,q\left(
t\right) \right) }{\partial q}\right) ^{2}\Psi  \notag \\
&&{-}\frac{\widetilde{\hslash }}{i}\frac{\delta }{\delta _{t}q\left(
t\right) }\int_{0}^{T}dt^{\prime }\delta \left( t-t^{\prime }\right) \frac{%
\partial \chi \left( t^{\prime },q\left( t^{\prime }\right) \right) }{%
\partial q}\Psi  \notag \\
&&{-}\left[ \left( \frac{\partial \chi \left( t,q\left( t\right) \right) }{%
\partial q}\right) ^{2}{+}\frac{\hslash }{i}\epsilon \delta \left( 0\right)
\frac{\partial ^{2}\chi \left( t^{\prime },q\left( t^{\prime }\right)
\right) }{\partial q^{2}}\right] \Psi .  \notag \\
\end{eqnarray}%
In the discrete approximation $\epsilon \delta \left( 0\right) =\epsilon
\left( 1/\epsilon \right) =1$, so the second variational derivative is
defined. The first term under the integral sign in Eq.~(\ref{8}) gives:
\begin{equation}
\overset{\cdot }{q}\frac{\widetilde{\hslash }}{i}\frac{\delta \Psi }{\delta
_{t}q\left( t\right) }=\overset{\cdot }{q}\frac{\partial \chi }{\partial q}%
\Psi =\left[ \frac{d\chi }{dt}-\frac{\partial \chi }{\partial t}\right] \Psi
.  \label{15}
\end{equation}%
Putting it all together, we obtain: the eigenvalue $\Lambda $ in Eq.(\ref{9}%
) does not depend on the internal points of the particle trajectory $q(t)$
and is equal to Eq.~(\ref{10}) if the wave function Eq.~(\ref{11}) obeys the
Schr\"{o}dinger equation Eq.~(\ref{3}).

The formulation of the QPLA for this simplest mechanical problem is easily
transferred to QTG. For a given initial state $\psi _{in}$, the evolution of
the universe is determined by the Schr\"{o}dinger equation
\begin{equation}
i\hslash \frac{\partial \psi }{\partial t}=\int_{\Sigma }d^{3}x\left( N%
\widehat{\mathit{H}}+N_{k}\widehat{\mathit{H}}^{k}\right) \psi ,  \label{16}
\end{equation}%
where $N,N_{k}$ are the Arnowitt, Deser, and Misner (ADM) lapse and shift
functions \cite{ADM}, and $\ \widehat{\mathit{H}}$,$\widehat{\mathit{H}}^{k}$%
are the constraint operators. We write the classical ADM constraints for the
simplest case, when matter consists of one real scalar field:
\begin{eqnarray}
\widehat{\mathit{H}} &\equiv &\frac{\kappa }{\sqrt{g^{\left( 3\right) }}}%
\left( Tr\mathbf{\pi }^{2}-\frac{1}{2}\left( Tr\mathbf{\pi }\right)
^{2}\right) -\frac{\sqrt{g^{\left( 3\right) }}}{\kappa }R^{\left( 3\right) }
\notag \\
&&-\sqrt{g^{\left( 3\right) }}\left[ \frac{p_{\varphi }^{2}}{2g^{\left(
3\right) }}+\frac{1}{2}g^{ik}\partial _{i}\varphi \partial _{k}\varphi
+V\left( \varphi \right) \right] .  \label{17}
\end{eqnarray}
\begin{equation}
\widehat{\mathit{H}}^{k}\equiv -2\pi _{\left. {}\right\vert
l}^{kl}-p_{\varphi }g^{kl}\partial _{l}\varphi .  \label{18}
\end{equation}%
Here $\pi ^{kl}$ and $p_{\varphi }$ are the momentum densities canonically
conjugate to the 3D metric $g_{kl}$ on the spatial section $\Sigma $ and the
scalar field $\varphi $, $\kappa =16\pi G(c=1)$. Canonical quantization is
reduced to replacing the canonical momenta with operators,

\begin{equation}
\widehat{\pi }^{kl}\left( x\right) \equiv \frac{\hslash }{i}\frac{\delta }{%
\delta g_{kl}\left( x\right) },\widehat{p}_{\varphi }\equiv \frac{\hslash }{i%
}\frac{\delta }{\delta \varphi \left( x\right) },  \label{19}
\end{equation}%
and substituting them into the gravitational constraints Eqs.~(\ref{17}) and
(\ref{18}). Here $\delta $ is the usual variation of the function of three
spatial variables. The problem of ordering non-commuting factors in Eq.~(\ref%
{16}) (see \cite{ChrZ}) is not discussed here.

The initial state of the universe $\psi _{in}$ is not, as we expect, a
solution of the WDW equation, and the wave function $\psi $ depends
explicitly on time. The kernel of the evolution operator for the Schr\"{o}%
dinger equation Eq.~(\ref{16}) can be represented by a functional integral
in which integration over the functions $N,N_{k}$ is not performed. This
evolution is unitary if the constraint operators in Eq.~(\ref{16}) are
Hermitian. In the transition to the QPLA formalism, the wave functional of
the universe is defined on the world history of $4D$ geometry and matter
fields: $\Psi \left[ g_{\mu \nu }\left( x^{\alpha }\right) ,\varphi \left(
x^{\alpha }\right) \right] $. Just like the classical action, the action
operator and its eigenvector $\Psi $ must be invariants of generally
covariant transformations. Arbitrary lapse and shift functions $N,N_{k}$
ensure this invariance in the same way as in classical gravity theory. Let
us represent the wave functional in exponential form:
\begin{eqnarray}
&&\Psi \left[ g_{\mu \nu }\left( x^{\alpha }\right) ,\varphi \left(
x^{\alpha }\right) \right]  \label{20} \\
&{=}&\exp \left\{ \frac{i}{\widetilde{\hslash }}\int_{0}^{T_{0}}dx^{0}\chi
^{0}\left( x^{0},g_{\mu \nu }\left( x^{0},\left[ x^{k}\right] \right)
,\varphi \left( x^{0},\left[ x^{k}\right] \right) \right) \right\} .  \notag
\end{eqnarray}%
The notations indicate that the fields enter $\chi ^{0}$ as functions of
time $x^{0}=t$, but $\chi ^{0}$ is a functional of the same fields as
functions of spatial coordinates. The function $\chi ^{0}$ is the logarithm
of the wave function of the universe -- the solution of the Schr\"{o}dinger
equation Eq.~(\ref{16}) in the direction of the time coordinate $x^{0}$.
Formula Eq.~(\ref{20}) is the first step towards a more symmetrical
representation of QTG with respect to space-time coordinates.

\section{GENERALIZED CANONICAL FORM OF ACTION OF THE THEORY OF GRAVITY}

In the standard canonical form of dynamical theory, the isolation of the
time parameter formally violates covariance already at the classical level.
However, there is a generalization of the canonical representation proposed
by De Donder \cite{DD1} and Weyl \cite{Weil} that eliminates this violation.
In this paper we will restrict our consideration to the simplest set of
matter fields consisting of one or more scalar fields, and therefore we will
begin our consideration with the action of a real scalar field in Lagrangian
form (metric signature $(+,-,-,-)$)
\begin{equation}
I_{\varphi }=\int \sqrt{-g}d^{4}x\left[ \frac{1}{2}g^{\mu \nu }\partial
_{\mu }\varphi \partial _{\nu }\varphi -V\left( \varphi \right) \right] .
\label{21}
\end{equation}%
Let us introduce the generalized canonical De Donder-Weil (DDW) momenta:
\begin{equation}
p_{\varphi }^{\mu }\equiv \frac{\partial \mathit{L}_{\varphi }}{\partial
\left( \partial _{\mu }\varphi \right) }=\sqrt{-g}g^{\mu \nu }\partial _{\nu
}\varphi ,  \label{22}
\end{equation}%
and also define the generalized density of the Hamiltonian function of a
scalar field:
\begin{equation}
\mathit{H}_{\varphi }=\partial _{\mu }\varphi p_{\varphi }^{\mu }-\mathit{L}%
_{\varphi }=\frac{1}{2}\frac{g_{\mu \nu }p_{\varphi }^{\mu }p_{\varphi
}^{\nu }}{\sqrt{-g}}+\sqrt{-g}V\left( \varphi \right) .  \label{23}
\end{equation}%
After this, we write the action Eq.~(\ref{21}) in the generalized canonical
form of the DDW:
\begin{equation}
I_{\varphi }=\int d^{4}x\left[ \partial _{\mu }\varphi p_{\varphi }^{\mu }-%
\mathit{H}_{\varphi }\right] .  \label{24}
\end{equation}

Let's do the same with the Hilbert-Einstein action,
\begin{equation}
I_{HE}=-\frac{1}{\kappa }\int d^{4}x\sqrt{-g}R.  \label{25}
\end{equation}%
Its Lagrangian density can be represented as \cite{LanLif}:
\begin{equation}
\kappa \mathit{L}_{HE}=-\sqrt{-g}R=\partial _{\mu }\left( \sqrt{-g}\Gamma
^{\mu }\right) -\sqrt{-g}G,  \label{26}
\end{equation}%
where
\begin{equation}
\Gamma ^{\mu }\equiv g^{\alpha \beta }\Gamma _{\alpha \beta }^{\mu }-g^{\mu
\alpha }\Gamma _{\alpha \beta }^{\beta },  \label{27}
\end{equation}
\begin{equation}
G\equiv g^{\mu \nu }\left( \Gamma _{\mu \beta }^{\alpha }\Gamma _{\nu \alpha
}^{\beta }-\Gamma _{\mu \nu }^{\alpha }\Gamma _{\alpha \beta }^{\beta
}\right) ,  \label{28}
\end{equation}%
and
\begin{equation}
\Gamma _{\left. a\right\vert bc}=\frac{1}{2}\left( \partial
_{b}g_{ac}+\partial _{c}g_{ab}-\partial _{a}g_{bc}\right)  \label{29}
\end{equation}%
are Christoffel symbols (the indices are raised and lowered using the metric
tensor). The first term on the right-hand side of Eq.~(\ref{26}) will be
important for formulating the boundary conditions at the corresponding
stages of the consideration. Now, to determine the generalized canonical
momenta of the gravitational field, it is convenient to start with the
momenta conjugate to the Christoffel symbols, which we define as follows:
\begin{equation*}
p^{\left. a\right\vert bc}=\frac{\partial \left( -\sqrt{-g}G/\kappa \right)
}{\partial \Gamma _{\left. a\right\vert bc}}
\end{equation*}
\begin{eqnarray}
&=&-\frac{\sqrt{-g}}{\kappa }\left[ \Gamma ^{\left. c\right\vert ba}+\Gamma
^{\left. b\right\vert ca}-g^{bc}\Gamma _{\left. {}\right. \left.
m\right\vert }^{\left. a\right\vert m}\right.  \notag \\
&&\left. -\frac{1}{2}\left( g^{ca}\Gamma _{\left. {}\right. \left. m\right.
}^{\left. b\right\vert m}+g^{ba}\Gamma _{\left. {}\right. \left. m\right.
}^{\left. c\right\vert m}\right) \right] .  \label{30}
\end{eqnarray}%
Just like $\Gamma _{\left. a\right\vert bc}$, the momenta $p^{\left.
a\right\vert bc}$ are symmetrical with respect to the pair of indices $bc$.
By folding equality Eq.~(\ref{30}) in two different ways, we find,
\begin{equation}
\frac{\sqrt{-g}}{\kappa }\Gamma _{\left. {}\right. \left. m\right\vert
}^{\left. a\right\vert m}=\frac{2}{3}p_{\left. {}\right. \left. m\right.
}^{\left. m\right\vert a},  \label{31}
\end{equation}
\begin{equation}
\frac{\sqrt{-g}}{\kappa }\Gamma _{\left. {}\right. \left. m\right. }^{\left.
m\right\vert a}=-\frac{1}{2}p_{\left. {}\right. \left. m\right. }^{\left.
a\right\vert m}+\frac{1}{3}p_{\left. {}\right. \left. m\right. }^{\left.
m\right\vert a}.  \label{32}
\end{equation}%
We also have:

\begin{eqnarray}
&&-\frac{\sqrt{-g}}{\kappa }\left( \Gamma ^{\left. c\right\vert ba}+\Gamma
^{\left. b\right\vert ca}\right)  \notag \\
&=&-\frac{\sqrt{-g}}{\kappa }\partial ^{a}g^{bc}=p^{\left. a\right\vert bc}+%
\frac{1}{2}g^{bc}p_{\left. {}\right. \left. m\right. }^{\left. a\right\vert
m}  \notag \\
&&+\frac{1}{3}\left( g^{ac}p_{\left. {}\right. \left. m\right. }^{\left.
m\right\vert b}+g^{ab}p_{\left. {}\right. \left. m\right. }^{\left.
m\right\vert c}-g^{bc}p_{\left. {}\right. \left. m\right. }^{\left.
m\right\vert a}\right) .  \label{33}
\end{eqnarray}%
Now we introduce the generalized canonical momenta of the DDW $\pi ^{\left.
a\right\vert bc}$ , conjugate to $\partial _{a}g_{bc}$, which we find from
the equality,
\begin{equation}
p^{\left. a\right\vert bc}\Gamma _{\left. a\right\vert bc}=\pi ^{\left.
a\right\vert bc}\partial _{a}g_{bc}.  \label{34}
\end{equation}%
We find:
\begin{equation}
\pi ^{\left. a\right\vert bc}=\frac{1}{2}\left( p^{\left. b\right\vert
ac}+p^{\left. c\right\vert ab}-p^{\left. a\right\vert bc}\right) ,
\label{35}
\end{equation}
\begin{equation}
p^{\left. c\right\vert ab}=\pi ^{\left. a\right\vert bc}+\pi ^{\left.
b\right\vert ca}.  \label{36}
\end{equation}%
After this, we write the Hilbert-Einstein action in the generalized
canonical form of the DDW:

\begin{equation}
I_{HE}=\int d^{4}x\left[ \partial _{a}g_{bc}\pi ^{\left. a\right\vert bc}-%
\mathit{H}_{g}\right] ,  \label{37}
\end{equation}%
where the density of the generalized Hamiltonian function is equal to

\begin{eqnarray}
&&\mathit{H}_{g}  \notag \\
&=&\frac{\kappa }{\sqrt{-g}}\left[ \pi _{\left. a\right\vert bc}\pi ^{\left.
b\right\vert ac}+\frac{1}{3}\pi _{a\left. {}\right. m}^{\left. {}\right\vert
m}\pi ^{\left. n\right\vert a}\left. _{n}\right. \right.  \notag \\
&&\left. +\frac{1}{3}\pi ^{\left. m\right\vert }\left. _{am}\right. \pi
^{\left. n\right\vert a}\left. _{n}\right. -\frac{1}{6}\pi _{a}^{\left.
{}\right\vert m}\left. _{m}\right. \pi ^{\left. a\right\vert n}\left.
_{n}\right. \right] .  \label{38}
\end{eqnarray}

\section{QUANTUM THEORY OF GRAVITY IN A GENERALIZED CANONICAL FORM}

We can solve the problem of quantizing the theory of gravity in a
generalized canonical form within the framework of the QPLA formalism. Let's
start with a real scalar field. The usual quantization procedure implies the
selection of the "time" parameter by $3+1$ partitioning of $4D$ space. In
the covariant theory, this parameter can be any coordinate $x^{a},a=1,2,3,4$%
. The three remaining coordinates will be denoted by $x^{\widetilde{a}}$ .
We will also introduce a rectangular lattice with elementary translation
vectors $\epsilon ^{a}$ in $4D$ space. We will consider the scalar field $%
\varphi \left( x^{a},\left[ x^{\widetilde{a}}\right] \right) $ as a
mechanical system with an infinite set of degrees of freedom, which are
numbered by the $3D$ index $x^{\widetilde{a}}$ , and moving in `time' $x^{a}$%
. We implement the components of the generalized operator of the momentum of
the scalar field on the space of wave functionals $\Psi \left[ \varphi
\left( x\right) \right] $ as follows. We represent the wave functional in
exponential form (there is no summation over the index $a$),
\begin{eqnarray}
&&\Psi \left[ \varphi \left( x\right) \right]  \notag \\
&=&\exp \left[ \frac{i}{\widetilde{\hslash }^{a}}\int_{0}^{T_{a}}dx^{a}\chi
^{a}\left( x^{a},\varphi \left( x^{a},\left[ x^{\widetilde{a}}\right]
\right) \right) \right] ,  \label{39}
\end{eqnarray}%
where
\begin{equation}
\widetilde{\hslash }^{a}=\hslash \epsilon ^{a}.  \label{40}
\end{equation}%
Then by definition:
\begin{equation}
\widehat{p}_{\varphi }^{a}\left( x\right) \Psi \equiv \frac{\delta \chi
^{a}\left( x^{a},\varphi \left( x^{a},\left[ x^{\widetilde{a}}\right]
\right) \right) }{\delta \varphi \left( x^{a},\left[ x^{\widetilde{a}}\right]
\right) }\Psi .  \label{41}
\end{equation}%
Let us emphasize once again that here the variational derivative is taken
with respect to the field $\varphi $ as a $3D$ functional in the space of
indices $x^{\widetilde{a}}$. Thus, we now have four functions $\chi
^{a}\left( x^{a},\varphi \left( x^{a},\left[ x^{\widetilde{a}}\right]
\right) \right) $, which represent one wave functional Eq.(\ref{39}) for
each possible choice of the time parameter. They are not independent, since
from Eq.~(\ref{39}) it follows:
\begin{equation}
\frac{1}{\epsilon ^{a}}\int_{0}^{T_{a}}dx^{a}\chi ^{a}=\frac{1}{\epsilon ^{b}%
}\int_{0}^{T_{b}}dx^{b}\chi ^{b},  \label{42}
\end{equation}%
where this relation remains finite in the continuum limit. We will also need
a representation of the square of the momentum operator. By analogy with
Eq.~(\ref{14}) we have:
\begin{eqnarray}
&&\widehat{p}_{\varphi }^{a}\left( x\right) \widehat{p}^{b}\left( x\right)
\Psi  \notag \\
&=&\frac{\widetilde{\hslash }^{a}}{i}\frac{\delta }{\delta \varphi \left(
x^{a},\left[ x^{\widetilde{a}}\right] \right) }\left[ \frac{\delta \chi
^{b}\left( x^{b},\varphi \left( x^{b},\left[ x^{\widetilde{b}}\right]
\right) \right) }{\delta \varphi \left( x^{b},\left[ x^{\widetilde{b}}\right]
\right) }\Psi \right]  \notag \\
&=&\left[ \frac{\delta \chi ^{a}\left( x^{a},\varphi \left( x^{a},\left[ x^{%
\widetilde{a}}\right] \right) \right) }{\delta \varphi \left( x^{a},\left[
x^{\widetilde{a}}\right] \right) }\frac{\delta \chi ^{b}\left( x^{b},\varphi
\left( x^{b},\left[ x^{\widetilde{b}}\right] \right) \right) }{\delta
\varphi \left( x^{b},\left[ x^{\widetilde{b}}\right] \right) }\right.  \notag
\\
&&\left. +\frac{\hslash }{i}\frac{\delta ^{2}\chi ^{b}\left( x^{b},\varphi
\left( x^{b},\left[ x^{\widetilde{b}}\right] \right) \right) }{\delta
\varphi \left( x^{a},\left[ x^{\widetilde{a}}\right] \right) \delta \varphi
\left( x^{b},\left[ x^{\widetilde{b}}\right] \right) }\right] \Psi .
\label{43}
\end{eqnarray}%
The singularity $\delta \left( 0\right) $ here arose (and was compensated by
multiplication by$\epsilon ^{a}$) in the second term due to the fact that
the first variational derivative $\delta \chi ^{b}\left( x^{b},\varphi
\left( x^{b},\left[ x^{\widetilde{b}}\right] \right) \right) /\delta \varphi
\left( x^{b},\left[ x^{\widetilde{b}}\right] \right) $ is a function of the
coordinate $x^{a}$, which is contained among $x^{\widetilde{b}}$ (if $a\neq
b $). The momentum operators of the gravitational field act in the same way.
Thus, the generalized Hamiltonian operators $\widehat{\mathit{H}}_{g}$ and $%
\widehat{\mathit{H}}_{\varphi }$ are defined (if we also agree to place the
momentum operators on the right in all terms). Now let us consider the terms
$\partial _{a}g_{bc}\widehat{\pi }^{\left. a\right\vert bc}$ and $\partial
_{a}\varphi \widehat{p}_{\varphi }^{a}$ under the action integral sign (here
the summation is performed over repeating indices):
\begin{eqnarray}
&&\left( \partial _{a}g_{bc}\widehat{\pi }^{\left. a\right\vert bc}+\partial
_{a}\varphi \widehat{p}_{\varphi }^{a}\right) \Psi  \notag \\
&=&\left[ \partial _{a}g_{bc}\frac{\delta \chi ^{a}\left( x^{a},g_{\alpha
\beta }\left( x^{a},\left[ x^{\widetilde{a}}\right] \right) ,\varphi \left(
x^{a},\left[ x^{\widetilde{a}}\right] \right) \right) }{\delta g_{bc}\left(
x^{a},\left[ x^{\widetilde{a}}\right] \right) }\right.  \notag \\
&&\left. +\partial _{a}\varphi \frac{\delta \chi ^{a}\left( x^{a},g_{\alpha
\beta }\left( x^{a},\left[ x^{\widetilde{a}}\right] \right) ,\varphi \left(
x^{a},\left[ x^{\widetilde{a}}\right] \right) \right) }{\delta \varphi
\left( x^{a},\left[ x^{\widetilde{a}}\right] \right) }\right] \Psi  \notag \\
&=&\left( d_{a}\chi ^{a}-\partial _{a}\chi ^{a}\right) \Psi ,  \label{44}
\end{eqnarray}%
where $d_{a}\chi ^{a}$ is the sum of the total derivatives of the components
of the wave function with respect to the `times' $x^{a}$, and $\partial
_{a}\chi ^{a}$ is the corresponding sum of the partial derivatives.

We are ready to formulate the QPLA in an arbitrary domain $\Omega $ and
derive the required consequences from it. The action operator has the form:
\begin{equation}
\widehat{I}_{\Omega }=\int d^{4}x\left[ \partial _{a}g_{bc}\widehat{\pi }%
^{\left. a\right\vert bc}+\partial _{a}\varphi \widehat{p}_{\varphi }^{a}-%
\widehat{\mathit{H}}_{g}-\widehat{\mathit{H}}_{\varphi }\right] .  \label{45}
\end{equation}%
The boundary contribution to the action will be considered separately in
each case. We are interested in the eigenvalue of the action operator, which
can be written in the form
\begin{equation}
\Lambda =\frac{\widehat{I}_{\Omega }\Psi }{\Psi }.  \label{46}
\end{equation}%
By definition, $\Lambda $ does not depend on the values of the metric tensor
and the matter fields inside the domain $\Omega $, but only on their
boundary values on $\partial \Omega $. The integral of the total derivative
in Eq.~(\ref{44}) gives the desired eigenvalue
\begin{equation}
\Lambda =\oint\limits_{\partial \Omega }dSn_{a}\chi ^{a},  \label{47}
\end{equation}%
$n_{a}$ is the unit vector of the outward normal to the surface $\partial
\Omega $. If the domain $\Omega $ is a rectangular parallelepiped, then the
integral Eq.~(\ref{47}) is reduced to the sum of the differences $\chi
^{a}\left( T_{a}\right) -\chi ^{a}\left( 0\right) $ on the opposite faces.
Thus, the eigenvalue of the action operator is reduced to the algebraic sum
(integral) of the boundary values of the components $\chi ^{a}$ of the wave
function, provided that the volumetric part of the action vanishes for any
configuration of geometry and matter fields inside $\Omega $. This condition
reduces to a variational-differential equation for the components of the
wave function:
\begin{equation}
\partial _{a}\chi ^{a}\Psi +\left( \widehat{\mathit{H}}_{g}+\widehat{\mathit{%
H}}_{\varphi }\right) \Psi =0.  \label{48}
\end{equation}%
The wave functional $\Psi $ in this equation is a common factor due to Eqs.~(%
\ref{41}) and (\ref{43}) and can be reduced. But we leave it, keeping in
mind some complication of the structure of the action in the observation
area. The nonlinear equation Eq.(\ref{48}) with respect to the components $%
\chi ^{a}$ of the wave function should be supplemented by proportions Eq.(%
\ref{42}). Equation Eq.~(\ref{48}) replaces the Schr\"{o}dinger equation Eq.(%
\ref{16}) in the new formulation of QTG.

\section{INITIAL STATE OF THE UNIVERSE}

Since we need time in QTG and the real dependence of the state of the
universe on time, we will also need the initial state of the universe. The
description of evolution in the formalism of QPLA is invariant with respect
to arbitrary transformations of space-time coordinates, if they do not
affect the initial and final spatial sections. Invariance is expressed in
the fact that the eigenvalue of the action operator (by definition) does not
depend on the internal points of the world history of geometry and matter
fields. In the work \cite{GLG2}, the QPLA in the generalized canonical form
of the DDW was used to formulate the quantum dynamics of the universe in the
`polar' region centered at the `South Pole' of the no-boundary
Hartle-Hawking wave function \cite{HH}. The boundary of the polar region --
the `polar circle' -- is, according to our assumption, a spatial section of
the universe $\Sigma _{in}$, beyond which evolution in real time begins. In
the polar region itself, the Euclidean form of the QPLA with imaginary time
is used. Thus, the difference between time and space coordinates completely
disappears, and there is no need to connect the singular point $g=\det
g_{\mu \nu }=0$ with the pole of the spherical coordinate system in which
the radii are the lines of time \cite{HH}. In this case, there is no need
for any boundary conditions at the pole. In order for such a formulation of
the QPLA to provide the necessary initial state of the universe $\psi _{in}$
on $\Sigma _{in}$, it is necessary to justify the existence of the
corresponding classical solution in the polar region -- the instanton. In
\cite{GLG1}, a justification was proposed for the $3D$ scale factor of the
universe $g^{\left( 3\right) }=\det g_{ij}$. It consists in the fact that
for the dynamics of this variable it is possible to prove an analogue of the
theorem of positivity of energy, based on the Witten identity \cite{Witt},
if the square of the eigenvalue of the $3D$ Dirac operator on a spatial
section is taken as the expansion energy. In this section we consider such a
justification for the dynamics of the conformal factor $\mathit{K}$ of the $%
4D$ metric, $g_{ab}=\mathit{K}^{2}B_{ab}$, where $B_{ab}$ is the base metric
with det$B_{ab}=1$\cite{Haw}. The Hilbert-Einstein Euclidean action in the
new variables takes the form:
\begin{equation}
I_{HE}=-\frac{1}{\kappa }\int_{\Omega }d^{4}x\left( \mathit{K}%
^{2}R_{B}+6B^{ab}\partial _{a}\mathit{K}\partial _{b}\mathit{K}\right) ,
\label{49}
\end{equation}%
where
\begin{equation}
R_{B}=\partial _{a}\left( B^{bc}\mathit{G}_{\left. {}\right. bc}^{a}\right)
+B^{ab}\mathit{G}_{\left. {}\right. ad}^{c}\mathit{G}_{\left. {}\right.
bc}^{d}  \label{50}
\end{equation}%
is the Ricci scalar, and $\mathit{G}_{\left. {}\right. bc}^{a}$ are the
Christoffel symbols for the base metric $B_{ab}$. We have not written out
the surface contributions to the action Eq.~(\ref{49}) (see \cite{Haw}),
which must be taken into account in the continuity conditions for the
components of the metric tensor $g_{ab}$ on the boundary $\partial \Omega
=\Sigma _{in}$. We are interested in the dynamics of the conformal factor $%
\mathit{K}$ in the domain $\Omega $. To simplify the consideration, we
choose harmonic coordinates in $\Omega $ \cite{DD}:
\begin{equation}
B^{bc}\mathit{G}_{\left. {}\right. bc}^{a}=0.  \label{51}
\end{equation}%
Now we represent the remaining expression in Eq.~(\ref{50}) as the
difference of two positive definite contributions:
\begin{eqnarray}
&&B^{ab}\mathit{G}_{\left. {}\right. ad}^{c}\mathit{G}_{\left. {}\right.
bc}^{d}  \notag \\
&=&B^{ab}B^{cp}B^{dq}\mathit{G}_{\left( p\right\vert \left. a\right) d}%
\mathit{G}_{\left( q\right\vert \left. b\right) c}  \notag \\
&&-B^{ab}B^{cp}B^{dq}\mathit{G}_{\left[ p\right\vert \left. a\right] d}%
\mathit{G}_{\left[ q\right\vert \left. b\right] c},  \label{52}
\end{eqnarray}%
where the round brackets denote symmetrization by a pair of indices, and the
square brackets denote antisymmetrization. The first of them corresponds to
the longitudinal components of the gravitational field (together with the
conformal factor), and the second to the transverse components associated
with gravitational waves. Thus, the density of the Lagrange function of the
theory of gravity after the selection of the conformal factor can be written
as:
\begin{eqnarray}
&&\mathit{L}_{g}  \notag \\
&=&-\frac{1}{\kappa }\left( 6B^{ab}\partial _{a}\mathit{K}\partial _{b}%
\mathit{K}+\mathit{K}^{2}B^{ab}B^{cp}B^{dq}\mathit{G}_{\left( p\right\vert
\left. a\right) d}\mathit{G}_{\left( q\right\vert \left. b\right) c}\right)
\notag \\
&&+\mathit{K}^{2}B^{ab}B^{cp}B^{dq}\mathit{G}_{\left[ p\right\vert \left. a%
\right] d}\mathit{G}_{\left[ q\right\vert \left. b\right] c}+\mathit{K}^{4}%
\mathit{L}_{m}.  \label{53}
\end{eqnarray}%
From the structure of the potential relief for the scale factor it is seen
that if the energy of the gravitational waves near the singularity $\mathit{K%
}=0$ is less than the longitudinal potential energy, classical motion of $%
\mathit{K}$ (in imaginary time) from the singular point is possible. This
means that there is an instanton in the polar region containing the
singularity $g=0$ inside.

On this basis, it can be expected that the QPLA in the generalized canonical
form of DDW,
\begin{equation}
\widehat{I}_{\Omega }\Psi =\Lambda _{\Omega }\Psi ,  \label{54}
\end{equation}%
has a non-trivial solution in the polar region. To enter the region of
motion with a real time parameter, we introduce spherical coordinates in the
polar region with a center at an arbitrary point (not associated with the
singularity). We pass in the action operator $\widehat{I}_{\Omega }$ to
spherical coordinates near the boundary of $\Omega $, where we write it in
the usual canonical form with the radial coordinate as the time parameter.
The wave functional $\Psi $ in these coordinates can be represented as a
product of wave functions defined on surfaces of constant radius. The
boundary wave function is equal to
\begin{equation}
\psi _{in}=\exp \left[ \frac{i}{\hslash }\Lambda _{\Omega }\right] ,
\label{55}
\end{equation}%
provided that $\partial \Omega =\Sigma _{in}$ is defined as the locus of the
cusp points of the scale factor:
\begin{equation}
\frac{\delta \Lambda _{\Omega }}{\delta \mathit{K}}=0.  \label{56}
\end{equation}%
At this boundary, a Wick rotation should be performed in the complex plane
of the radial coordinate. We emphasize once again that the initial state of
the universe determined in this way is not a solution of the WDW equation,
and the proper mass of the universe in this state is not equal to zero.

\section{THE DOMAIN OF OBSERVATION IN THE UNIVERSE}

Let us return to the space-time with the Lorentz signature and select some
compact, convex region $\Omega $ in it. We modify the quantum theory here by
the additional condition Eq.(\ref{1}). If we start with the functional
integral representing the evolution operator for the Schr\"{o}dinger
equation Eq.~(\ref{16}), the modification can be achieved by limiting the
measure of the functional integration by the factor \cite{Mens}
\begin{equation}
\delta ^{\Omega }\left( \nabla _{\mu }T^{\mu \nu }\right) ,  \label{57}
\end{equation}%
where $\delta ^{\Omega }$ is the functional delta function with support $%
\Omega $. This is equivalent to adding to the classical action of the
integral
\begin{equation}
-\int_{\Omega }\sqrt{-g}d^{4}x\lambda _{\nu }\nabla _{\mu }T^{\mu \nu }
\label{58}
\end{equation}%
with an arbitrary vector Lagrange multiplier $\lambda _{\nu }$. After
integration by parts,
\begin{equation}
\int_{\Omega }\sqrt{-g}d^{4}x\nabla _{\mu }\lambda _{\nu }T^{\mu \nu
}-\oint\limits_{\partial \Omega }dS\sqrt{-g}n_{\mu }\lambda _{\nu }T^{\mu
\nu },  \label{59}
\end{equation}%
the Lagrange multiplier will be included in the dynamic variables. The
surface integral over the boundary of the observation region will be
compensated after the main part of the action is reduced to a generalized
canonical form. Thus, the density of the Lagrange function of the modified
theory has the form:
\begin{equation}
\mathit{L}=\mathit{L}_{g}+\mathit{L}_{\varphi }+\sqrt{-g}\nabla _{\mu
}\lambda _{\nu }T^{\mu \nu },  \label{60}
\end{equation}%
where
\begin{equation}
\mathit{L}_{\varphi }=\sqrt{-g}\left[ \frac{1}{2}g^{\mu \nu }\partial _{\mu
}\varphi _{k}\partial _{\nu }\varphi _{k}-V\left( \varphi \right) \right] ,
\label{61}
\end{equation}%
is the density of the Lagrange function of the multiplet of scalar fields of
matter, and
\begin{equation}
T^{\mu \nu }=g^{\mu \alpha }g^{\nu \beta }\partial _{\alpha }\varphi
_{k}\partial _{\beta }\varphi _{k}-\frac{g^{\mu \nu }\mathit{L}_{\varphi }}{%
\sqrt{-g}}  \label{62}
\end{equation}%
is the energy-momentum tensor of matter fields. For greater generality, we
consider here a multiplet of scalar fields. We define the generalized
canonical momenta of the DDW. The generalized momenta for the Lagrange
multipliers $\lambda _{\nu }$ are equal to the density of the
energy-momentum tensor:
\begin{equation}
\mathit{P}^{\mu \nu }=\sqrt{-g}T^{\mu \nu }.  \label{63}
\end{equation}%
The generalized momenta associated with the Christoffel symbols and matter
fields will be modified by adding new terms:
\begin{equation}
\widetilde{p}^{\left. a\right\vert bc}=p^{\left. a\right\vert bc}-\lambda
^{a}\mathit{P}^{bc},  \label{64}
\end{equation}
\begin{equation}
\widetilde{p}_{k}^{\mu }=\sqrt{-g}\left[ g^{\mu \nu }+2\nabla _{\left(
a\right. }\lambda _{\left. b\right) }\mathit{T}^{ab\left\vert \mu \nu
\right. }\right] \partial _{\nu }\varphi _{k},  \label{65}
\end{equation}%
where indicated
\begin{equation}
\mathit{T}^{ab\left\vert \mu \nu \right. }=g^{\mu \alpha }g^{\nu b}-\frac{1}{%
2}g^{\mu \nu }g^{ab}.  \label{66}
\end{equation}%
In order to express, as usual, the generalized velocities $\partial _{\nu
}\varphi _{k}$ in terms of the generalized momenta $\widetilde{p}_{k}^{\mu }$%
, we must use relations Eq.~(\ref{65}), where the square bracket contains a
symmetric, second-rank tensor $\nabla _{\left( a\right. }\lambda _{\left.
b\right) }$, which is composed of derivatives and must also be excluded. We
will proceed differently. Let us first consider the simplest case, when
there is a single real scalar field of matter. The combination of equations
Eq.~(\ref{61}), Eq.~(\ref{62}) and Eq.(\ref{63}) then gives:

\begin{equation}
\mathit{T}_{\mu \nu \left\vert ab\right. }^{-1}\left( \frac{\mathit{P}^{ab}}{%
\sqrt{-g}}-g^{ab}V\right) =\partial _{\mu }\varphi \partial _{\nu }\varphi ,
\label{67}
\end{equation}%
where $\mathit{T}_{\mu \nu \left\vert ab\right. }^{-1}$ is a matrix with
paired indices, inverse to Eq.(\ref{66}). From here we find:

\begin{equation}
\partial _{\mu }\varphi =\sqrt{\mathit{T}_{\mu \mu \left\vert ab\right.
}^{-1}\left( \frac{\mathit{P}^{ab}}{\sqrt{-g}}-g^{ab}V\right) }.  \label{68}
\end{equation}%
For given $\mathit{P}^{ab}(x)$ and $g^{ab}(x)$ these differential equations
are first integrals of the classical equation of motion of a scalar field
for each possible choice of the "time" parameter. Integrating further Eq.(%
\ref{68}), we find the increments of coordinates $\Delta x^{\mu }$ between
the cusp points $\partial _{\mu }\varphi =0$, measured by the dynamics of
the scalar field in the corresponding coordinate direction.

Let's complicate the problem: let there be a triplet of scalar fields $%
\varphi _{k},k=1,2,3$. Equation (\ref{67}) will be replaced by the
following:
\begin{equation}
\mathit{T}_{\mu \nu \left\vert ab\right. }^{-1}\left( \frac{\mathit{P}^{ab}}{%
\sqrt{-g}}-g^{ab}V\right) =\partial _{\mu }\varphi _{k}\partial _{\nu
}\varphi _{k},  \label{69}
\end{equation}%
where the summation is performed over the index $k$. In this case, we also
extract the square root using the Dirac method \cite{Dir}. We replace the
triplet $\varphi _{k}$ with a $2\times 2$ matrix,

\begin{equation}
\widehat{\varphi }=\varphi _{k}\widehat{\sigma }^{k},  \label{70}
\end{equation}%
where $\widehat{\sigma }^{k}$ are the Pauli matrices, and we write down the
formal solution of equations (\ref{69}):

\begin{equation}
\partial _{\mu }\widehat{\varphi }=\sqrt{\mathit{T}_{\mu \mu \left\vert
ab\right. }^{-1}\left( \frac{\mathit{P}^{ab}}{\sqrt{-g}}-g^{ab}V\right) },
\label{71}
\end{equation}%
where the expression on the right is multiplied by the unite $2\times 2$
matrix. These matrix equations will acquire a certain meaning below after
introducing the components of the wave function. By collapsing both sides of
Eq.~(\ref{65}) with the Pauli matrices, we also write it in matrix form:
\begin{equation}
\widehat{\widetilde{p}}_{k}^{\mu }=\sqrt{-g}\left[ g^{\mu \nu }+2\nabla
_{\left( a\right. }\lambda _{\left. b\right) }\mathit{T}^{ab\left\vert \mu
\nu \right. }\right] \partial _{\nu }\widehat{\varphi }.  \label{72}
\end{equation}%
Substituting Eq.~(\ref{71}) into Eq.~(\ref{72}), we obtain a matrix equation
in which the components of the symmetric tensor $\nabla _{\left( a\right.
}\lambda _{\left. b\right) }$ remain unknown. These equations are sufficient
to determine the components of $\nabla _{\left( a\right. }\lambda _{\left.
b\right) }$ as a linear function of the generalized momenta $\widehat{%
\widetilde{p}}_{k}^{\mu }$ in matrix form. Thus, $\nabla _{\left( a\right.
}\lambda _{\left. b\right) }$ should now also be considered as $2\times 2$
matrices.

Completing the modification of quantum theory in the observation domain $%
\Omega $, we find the generalized canonical form of action of the modified
theory,
\begin{eqnarray}
\widetilde{I}_{\Omega _{obs}} &=&\int_{\Omega _{obs}}d^{4}x\left[ \pi
^{\left. a\right\vert bc}\partial _{a}g_{bc}+p_{k}^{\mu }\partial _{\mu
}\varphi _{k}\right.  \notag \\
&&\left. +\mathit{P}^{ab}\nabla _{\left( a\right. }\lambda _{\left. b\right)
}-\widetilde{\mathit{H}}\right] ,  \label{73}
\end{eqnarray}%
where the density of modified Hamiltonian is equal:
\begin{eqnarray}
\widetilde{\mathit{H}} &=&\mathit{H}_{g}+\sqrt{\mathit{T}_{\mu \mu
\left\vert ab\right. }^{-1}\left( \frac{\mathit{P}^{ab}}{\sqrt{-g}}%
-g^{ab}V\right) }\widehat{p}^{\mu }  \notag \\
&&-\frac{1}{2}g^{\mu \nu }\mathit{T}_{\mu \nu \left\vert ab\right. }^{-1}%
\mathit{P}^{ab}+3\sqrt{-g}V.  \label{74}
\end{eqnarray}%
$\mathit{H}_{g}$ is the density of the generalized Hamiltonian function of
the original theory Eq.~(\ref{38}). In the final expressions, we have
discarded the now unnecessary "tilde" signs at the canonical momenta. Thus,
the gravitational part of the action in the observation region is not
changed. However, there are two significant features that require separate
consideration.

The first feature is the appearance of a term
\begin{eqnarray}
&&\int_{\Omega _{obs}}d^{4}x\mathit{P}^{ab}\nabla _{\left( a\right. }\lambda
_{\left. b\right) }  \notag \\
&=&-\int_{\Omega _{obs}}d^{4}x\nabla _{a}\mathit{P}^{ab}\lambda
_{b}+\oint\limits_{\partial \Omega _{obs}}dSn_{a}\lambda _{b}\mathit{P}^{ab}
\label{75}
\end{eqnarray}%
in the action. After integration by parts, $\lambda _{\nu }$ play the role
of the usual canonical momenta, to which the four components of $\mathit{P}%
^{\nu a}$ are conjugate, if the coordinate $x^{a}$ is chosen as the time
parameter. Since $\lambda _{\nu }$ is not contained in the modified
Hamiltonian function Eq.(\ref{74}), we obtain the classical continuity
equations for the energy-momentum density of matter
\begin{equation}
\nabla _{\mu }\mathit{P}^{\mu a}=0,  \label{76}
\end{equation}%
which can be solved before quantizing the modified theory. This can be done
if the conditions on the boundary of the observation region are specified.
We have two boundary contributions: one in formula Eq.~(\ref{75}), and the
other in formula Eq.~(\ref{59}). They cancel each other out due to equality
Eq.~(\ref{63}). Thus, there are no contributions to the action on the
boundary of the observation region, but the question of fixing the boundary
conditions for Eq.~(\ref{76}) remains open.

The second feature of the modified action is its linear dependence on the
canonical momenta of the matter fields $\widehat{p}^{\mu }$ and the matrix
nature of this dependence. In the simplest case, when there is only one
scalar field, this means that in addition to Eq.~(\ref{76}), we also have
four equations of the form Eq.~(\ref{68}), determining the geometric
parameters $\Delta x^{\mu }$ of the observation region according to the
classical dynamics of this field. These equations can also be solved before
quantization. If there are more matter fields, after quantization we obtain
a spinor equation for the components of the wave functional (summation is
performed over the index $\mu $),
\begin{equation}
\partial _{\mu }\varphi _{k}\epsilon ^{\mu }\frac{\delta \widehat{\Psi }%
_{\mu }}{\delta _{\mu }\varphi _{k}}=\sqrt{\mathit{T}_{\mu \mu \left\vert
ab\right. }^{-1}\left( \frac{\mathit{P}^{ab}}{\sqrt{-g}}-g^{ab}V\right) }%
\widehat{\sigma }^{k}\epsilon ^{\mu }\frac{\delta \widehat{\Psi }_{\mu }}{%
\delta _{\mu }\varphi _{k}},  \label{77}
\end{equation}%
which is now represented by four products of $2\times 2$ martix wave
functions:
\begin{equation}
\widehat{\Psi }_{\mu }=\widehat{T}_{\mu }\prod\limits_{n_{\mu }}\widehat{%
\psi }^{\mu }\left( n_{\mu }\epsilon ^{\mu },q\left( n_{\mu }\epsilon ^{\mu
},\left[ x^{\widetilde{\mu }}\right] \right) \right)  \label{78}
\end{equation}%
for each choice of the time parameter $x^{\mu }$. Here $n_{\mu
}=1,2,...,N_{\mu },T_{\mu }=N_{\mu }\epsilon ^{\mu }$, $q=\left[ g_{\mu \nu
},\varphi _{k},\mathit{P}^{ab}\right] $. The products in Eq.~(\ref{78}) are
chronologically ordered in the corresponding `time'. We still assume that
Eq.~(\ref{78}) gives different multiplicative representations of the same
functional $\widehat{\Psi }_{\Omega _{obs}}$, which is a $2\times 2$ matrix.
Such a complication of quantum theory in the observation region raises the
question of consistency at the boundary $\partial \Omega _{obs}$ with the
rest of the universe. Just as on the boundary with the polar region, here we
need to set the boundary wave function of the universe. And we will do the
same here: if $\Lambda _{obs}$ is the eigenvalue of the action operator $%
\widehat{\widetilde{I}}_{\Omega _{obs}}$ corresponding to the eigenvector $%
\widehat{\Psi }_{\Omega _{obs}}$ in the observation region, then as the
boundary value of the wave function we will take
\begin{equation}
\psi _{\partial \Omega _{obs}}=\exp \left[ \frac{i}{\hslash }\Lambda _{obs}%
\right] .  \label{79}
\end{equation}

\section{OBSERVER IN QUANTUYM COSMOLOGY}

We have spoken about the region of observation with special laws of motion
of matter Eqs.~(\ref{76}) and (\ref{77}) inside. Let us now call this region
the `observer' and formulate the consequences that can be obtained from the
QPLA for its cosmological evolution. The operator of action of the universe
is additive and consists of two parts:
\begin{equation}
\widehat{I}=\widehat{\widetilde{I}}_{\Omega _{obs}}+\widehat{I}_{\Omega
^{\prime }},  \label{80}
\end{equation}%
where the first term is obtained by quantizing the modified action Eq.~(\ref%
{73}), and the second is the quantized action Eq.(\ref{45}) for the rest of
the universe. The first term in Eq.~(\ref{80}) describes the motion of the
energy-momentum $\mathit{P}^{ab}$ inside the observer, as well as the
geometry of its boundary $\Delta x^{\mu }$, and the full action describes
the dynamics of the fundamental degrees of freedom in the entire universe.
All this is contained in the secular equation
\begin{equation}
\left( \widehat{\widetilde{I}}_{\Omega _{obs}}+\widehat{I}_{\Omega ^{\prime
}}\right) \Psi =\Lambda \Psi ,  \label{81}
\end{equation}%
where $\Psi =\widehat{\Psi }_{\Omega _{obs}}\Psi _{\Omega ^{\prime }}.$The
eigenvalue $\Lambda $, by definition, is a $c$ -number that depends only on
the boundary values of the dynamic variables. What is important to us is
what happens on the inner boundary $\partial \Omega _{obs}$, separating the
observer and the rest of the universe. On it, the fundamental degrees of
freedom must be continuous. In addition, to exclude the boundary
contribution arising from the total derivative in Eq.~(\ref{26}), the normal
components Eq.~(\ref{27}) must also be continuous. Earlier, we also noted
the equality of the normal components $n_{\mu }\mathit{P}^{\mu a}$ and $%
\sqrt{-g}n_{\mu }T^{\mu a}$ on the boundary, which is also satisfied by Eq.~(%
\ref{63}). If all this is satisfied, $\Lambda $ `does not feel' what is
happening on the boundary $\partial \Omega _{obs}$. However, if we fix some
admissible values of the normal components $n_{\mu }\mathit{P}^{\mu a}$ on
the boundary, we constrain the set of trajectories in the configuration
space of the universe, on which, by definition, $\Lambda $ does not depend.
And $\Lambda $ will `feel' these admissible values. What determines the
admissible values of $n_{\mu }\mathit{P}^{\mu a}$? If on earlier sections of
the boundary $\partial \Omega _{obs}$ the energy-momentum flux is determined
only by the previous state of the universe, on later sections the continuity
equations Eq.~(\ref{76}) inside the observer must be taken into account. In
other words, the internal dynamics of the observer imposes restrictions on
the admissible boundary values $n_{\mu }\mathit{P}^{\mu a}$, and this is
`felt' by $\Lambda $.

Thus, the formulation of the QTG in terms of the modified QPLA Eq.~(\ref{81}%
) allows us to introduce into consideration the cosmological mechanism of
decoherence, in which a certain state of the observer $\widehat{\Psi }%
_{\Omega _{obs}}$ is also accompanied by a certain final state of the
universe
\begin{equation}
\psi _{out}=\exp \left[ \frac{i}{\hslash }\Lambda _{out}\left( n_{\mu }%
\mathit{P}^{\mu a}\right) \right] .  \label{82}
\end{equation}

The direct mathematical connection between what the observer `sees' ($n_{\mu
}\mathit{P}^{\mu a}$ on $\partial \Omega _{obs}$) and the future state of
the universe $\psi _{out}$ can be seen as a justification for the
many-worlds interpretation of Everett's quantum mechanics. At the same time,
the final state $\psi _{out}$ also has a numerical characteristic: its norm
depends on the result of local observation. Indeed, the restriction of the
set of admissible trajectories in defining $\Lambda $ in Eq.~(\ref{81}),
which is the essence of cosmological decoherence, is equivalent to the
restriction of the integration measure in the functional integral. The
latter obviously entails a violation of the unitarity of the evolution
operator. The presence of such a numerical characteristic of the final state
of the universe as $\left\vert \left\vert \psi _{out}\left( n_{\mu }\mathit{P%
}^{\mu a}\right) \right\vert \right\vert $ leaves some ambiguity in the
picture of the `branching' of the universe. If we accept the simultaneous
realization of all possible observation results and the corresponding
branches of cosmological evolution, how should we interpret the mentioned
numbers corresponding to each branch? The probabilistic interpretation does
not fit into this picture. To interpret the numerical parameterization of
the various branches of the Everett multiverse, we will choose a more
suitable one, which is closer to the original Hilbert-Einstein principle of
least action: the eigenvalue of the action operator of the universe at the
final stage of its evolution $\Lambda _{out}\left( n_{\mu }\mathit{P}^{\mu
a}\right) $. More precisely, we need its real part, averaged over the final
state (77):

\begin{equation}
S_{obs}\left[ n_{\mu }\mathit{P}^{\mu a}\right] =\left\langle Re%
(\Lambda _{out}\left( n_{\mu }\mathit{P}^{\mu a}\right)) \right\rangle _{\psi
_{out}}.  \label{83}
\end{equation}

\section{CONCLUSIONS}

The modification of the QPLA, by separating the observation region whose
boundary is determined by the internal deterministic dynamics of energy and
momentum using equations Eqs.~(\ref{76}) and (\ref{77}), also assumes
certain boundary conditions for this internal dynamics. This deterministic
dynamics, introduced as an additional condition in quantum theory, serves as
a mechanism of decoherence. The cosmological aspect of decoherence is that
the observation domain violates unitarity and reduces cosmological evolution
to different states corresponding to the observer's state. In this we see
the justification for Everett's picture of the multiverse. However, the
dependence of the final state of the universe on the state of the observer
obtained in this work allows us to interpret this multiverse as a set of
trial states in the cosmological principle of least action. It is the only
state (if such exists) that corresponds to the extremum of the norm that is
actualized. In this case, the final state norm should be understood as an
action functional in quantum cosmology, the extremum of which determines the
observed world history of the universe, as seen by the observer. The
variational parameters here are the energy and momentum flows $n_{\mu }%
\mathit{P}^{\mu a}$ (as well as other observables) on the boundary $\partial
\Omega _{obs}$.

In this paper we limited ourselves to a detailed consideration of the
covariant law of conservation of the energy-momentum tensor Eq.~(\ref{1}) in
the observation region. However, other Noether identities should be included
in the additional conditions of the modified QTG. In particular, the law of
conservation of electric charge in the observation area must be taken into
account. No fundamental complications, such as the appearance of spinor
variables, are expected here.

\section{ACKNOWLEDGEMENTS}

We are thanks V.A. Franke for useful discussions.




\end{document}